\documentclass[%
aip,
rsi,%
amsmath,amssymb,
reprint,%
]{revtex4-1}

\usepackage{graphicx}% Include figure files
\usepackage{dcolumn}% Align table columns on decimal point
\usepackage{bm}% bold math
\usepackage{algorithm}%
\usepackage{algorithmicx}%
\usepackage{algpseudocode}%
\usepackage{float}
\usepackage[colorlinks=true, linkcolor=blue, citecolor=blue, urlcolor=blue]{hyperref}

\begin{document}
	
	\title{The revival of holographic photonic integration }[HOLOGRAPHIC WAVEGUIDES]
	
\author{Daniel Brunner}
\address{FEMTO-ST Institute/Optics Department, CNRS - University Franche-Comté, 15B avenue des Montboucons, Besançon Cedex, 25030, France}
\email{daniel.brunner@femto-st.fr}

	\date{\today}% It is always \today, today,
	
%	\begin{abstract}
		
%In recent years, the hardware implementation of neural networks, leveraging physical coupling and analog neurons has substantially increased in relevance. Such nonlinear and complex physical networks provide significant advantages in speed and energy efficiency, but are potentially susceptible to internal noise when compared to digital emulations of such networks. In this work, we consider how additive and multiplicative Gaussian white noise on the neuronal level can affect the accuracy of the network when applied for specific tasks and including a softmax function in the readout layer. We adapt several noise reduction techniques to the essential setting of classification tasks, which represent a large fraction of neural network computing. We find that these adjusted concepts are highly effective in mitigating the detrimental impact of noise.	
		
%	\end{abstract}
	
	\maketitle
	
%\section{Introduction}

Photonic integration of thick holograms in wave-guiding structures could be considered the chimera of photonics; multi-faceted and hard to tame. It is the fundamental, and hence indispensable, concept behind compact and monolithically integrated linear optical transformation \cite{kip2006photorefractive}. The true relevance of this becomes apparent in the high-dimensional context of unconventional optical computing, that is, in optical neural networks. Yet, integrating such holographic connections is very challenging. It demands high fabrication accuracy, and numerical design of the circuit is often non-tractable for large architectures. Both challenges are intrinsically linked to the usually large refractive index differences between sections of such holographic optical waveguides when using standard techniques of silicon photonics. 

\begin{figure}[t]
	\centering
	\includegraphics[width=1\linewidth]{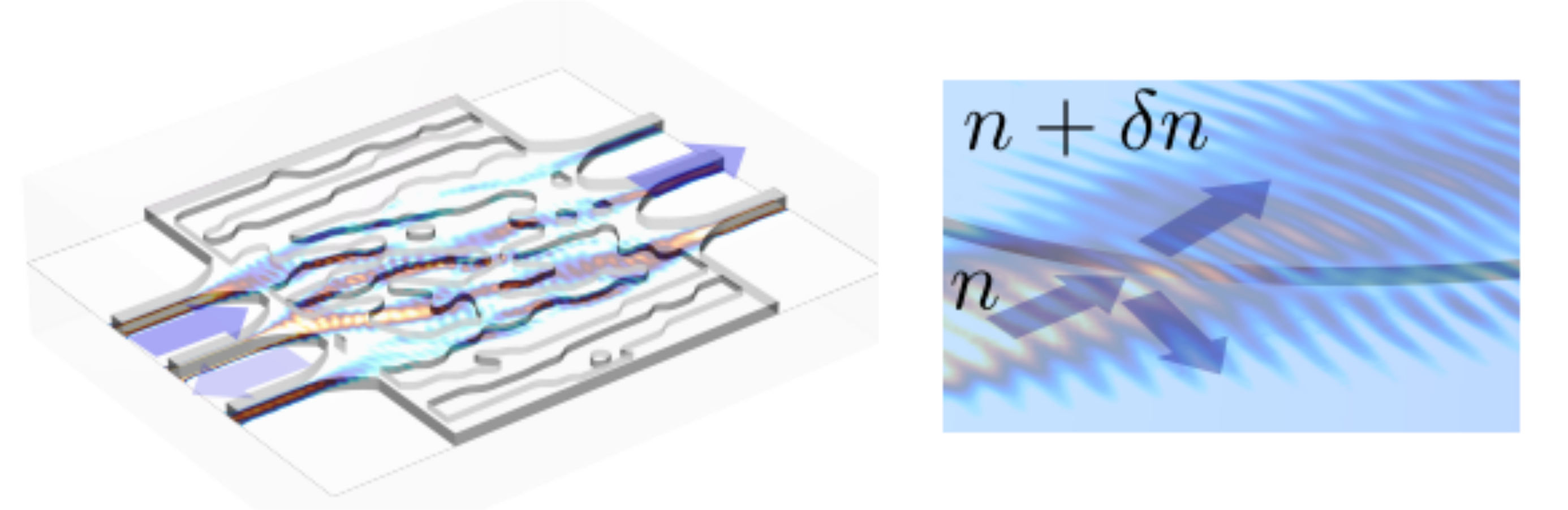}  % Adjust width as needed
	\caption{\bf{Schematic of the inverse-designed structures in Si-photonic platform reported in ref. cite{nikkhah2024inverse}}. When a guided optical wave passes an interface with a significant discontinuity in height and refractive index, see a, it experiences back reflection and potentially out of plane scattering. This creates resonances, increases losses and necessitates more complex, three-dimensional numerical modelling. b, Integrating such structures using more modest index modifications reduces this effect, making numerical inverse design computational tractable for large structures while enabling fabrication via commercial foundry processes. Figure adapted from ref. \cite{nikkhah2024inverse}.}
	\label{fig:Schematic}
\end{figure}

Reporting in Nature Photonics, Vahid Nikkhah and colleagues\cite{nikkhah2024inverse} introduce and experimentally demonstrate a versatile concept of moderate refractive-index modulated holographic waveguides based on shallow etching guiding structures integrated in a silicon-on-insulator (SOI) platform. The team use a standard option in a commercially available foundry process, making the concept rapidly and widely employable. 
The common SOI integration approach with commercial processes creates binary structures with large refractive index modifications for optical confinement or complex diffraction at non-trivial interfaces, see Fig.~\ref{fig:Schematic}(a). Such high refractive index contrast of usually $\Delta\text{n}\geq 1$ is a great asset for high-density integration leveraging single-mode waveguide circuitry. However, when it comes to complex multi-mode circuits that are not pre-mediated and, for example, designed by flexible inverse design concepts, this large $\Delta\text{n}$ can convert into a challenge. Optical electro-magnetic waves crossing an interface with such a large $\Delta\text{n}$ and a discontinuity in height are partially back-reflected, as well as often scattered out of plane. The first effect results in potentially strong optical resonances, limiting the spectral window in which the device can be employed. The second effect potentially necessitates expanding numerical simulations into the third-dimension perpendicular to the waveguide, which quickly makes numerical simulations of large circuits too cumbersome for iterative optimization in an inverse design approach.

The mathematical equivalent of such circuits is the multiplication of complex matrix T onto a complex valued input vector encoded in the circuits input field. As such, they optically implement linear complex valued vector matrix multiplication, just as the now widely employed meshes of discrete integrated interferometers \cite{bogaerts2020programmable}. These linear operations are the bases of numerous applications, for example, in imaging tasks where they can realize convolutions of an input according to some kernel, or invert the scattering, in order to create a clear image when imaging through some fixed complex media. However, and most relevant in the context of recent activity and interest in the photonics community, such devices that can implement the linear vector matrix products are so costly in neural network computing when using classical computers \cite{shen2017deep}.
 
Changing from deep to shallow etching allows an almost continuous tuning of $\Delta\text{n}$ without the addition of further materials. Using the example of Nikkhah et al. \cite{nikkhah2024inverse}, a 220-nm-thick silicon (Si) slab embedded in silicon oxide (SiO2) results in $n_{\text{eff,220}}=2.86$ as the effective refractive index of its fundamental mode. Reducing the Si layer thickness down to 150 nm through a shallow etch slightly reduces this to $n_{\text{eff,150}}=2.56$. Creating refractive index modifications of $\Delta\text{n}=0.3$ through locally etching the wide Si slab, Nikkhah and colleagues implement the non-trivial and inverse-design optimized refractive index profile of a holographic waveguide with strongly reduced back-reflections at interfaces. Importantly, the effect experienced by a propagating wave at the etching transitions’ height discontinuity is well approximated by the two-dimensional (2D) model underlying their iterative circuit optimization. For that the modes before and after the interface need to significantly overlap, which again is ensured by the similarity of their neff and the smaller reduction in height through the shallow etch. Under such assumptions, Nikkhah and colleagues introduce a model referred to as the propagation-based 2D Effective Index Approximation (p2DEIA), which captures the propagation within a uniformly etched region, angle of refraction and angle of reflection within the limits imposed by the demonstrated circuit complexity2. 
The team use p2DEIA to inversely design various holographic waveguides to implement predetermined complex-valued transmission matrices T. They find that at the design wavelength, the p2DEIA algorithm well approximates the target matrix as well as the full 3D simulation for a 2x2 and 3x3 transmission matrix T. The concepts validity is then demonstrated by experimentally implementing such multiplications for two- and three-dimensional input vectors, and the fabricated circuit does implement the desired transformation. According to them, when expanding the hologram’s size to 10x10, a full 3D simulation already becomes too demanding for the optimization, yet leveraging p2DEIA, they could fine tune their design during hundreds of iterations. The optimization successfully converges, however they point out that for structures of such and larger sizes (here more than 50x50 times the optical wavelength), the numerical model for optical effects at the interfaces should be improved. 

Due to their application and technological relevance, such holographic waveguides have been discussed in the last 25 years. Two interesting options arise when creating their refractive index distribution: a permanent and pre-designed structuring, or dynamical refractive index adjustment leveraging some additional, usually also optical, control mechanism. Mostly the first has already been reviewed in ref. \cite{kip2006photorefractive}, where however most techniques resulted in a very small and often temporally not stable modification of the refractive index distribution. A more versatile approach, in particular in the context of neural networks or generally unconventional computing, has been demonstrated in ref. \cite{brady1991holographic}, where holographic diffraction in a large 2D guiding slab of a photorefractive material was optically created by means of illuminating an optical control pattern on the waveguide slab from the top. This concept has recently experienced a strong revival motivated by the rise of optical neural networks and their in most cases equally iterative, hence dynamical, optimization during a learning stage. Improving versatility, a lithium niobate waveguide was sandwiched between a Si substrate and a photoconductor layer cite{onodera2024scaling}. Illumination from the top programs the spatial distribution of the voltage applied across the lithium niobate waveguide, and hence, as in ref. \cite{brady1991holographic}, allows the dynamic tuning of the holographic medium. Finally, the general idea of such complex multi-mode waveguides has been extended recently to optically control the spatial distribution of the imaginary index inside a guiding semiconductor structure \cite{wu2023lithography}, and hence optical gain is now available for such concepts. Additionally, such holographic waveguides can now be extended to 3D photonic integration \cite{porte2021direct}, where single-step additive fabrication allows combining optical waveguides and holographic refractive index distributions in a (3+1)D context in order to achieve large-scale vector matrix multiplication in integrated photonics.

Compared with the above-mentioned previous works and the other recent activities on holographic waveguide integration, the work by Nikkhah et al. \cite{nikkhah2024inverse} extends the toolbox towards commercial and hence economic fabrication. It remains to be seen if their approach allows to implement large-scale linear vector matrix multiplication with the fidelity required by pre-designed neural networks and other computing concepts. However, it has significant potential for implementing such operations on a more medium scale, for which the numerical optimization concept based on p2DEIA would most likely have to be further improved in accuracy to reach between 10 and 100 input dimensions. Due to its entirely passive nature, such chips do promise great energy efficiency and could play an important role in distributed sensing applications and edge computing.

\begin{acknowledgments}
	
	The authors declare no competing interests..

\end{acknowledgments}

\section*{References}
\bibliography{bibliography}% Produces the bibliography via BibTeX.
	
\end{document}